\documentclass[aps,pre,reprint,groupedaddress]{revtex4-1}

\usepackage{graphicx}
\usepackage{amssymb, amsmath}
\usepackage[caption=false]{subfig}
\usepackage{xcolor}
\usepackage{float}

\begin{document}

%\preprint{}

\title{Hyperuniformity Order Metric of Barlow Packings}


\author{T. M. Middlemas}
%\thanks{}
\affiliation{Department of Chemistry, Princeton University, New Jersey 08544, USA}

\author{F. H. Stillinger}
\affiliation{Department of Chemistry, Princeton University, New Jersey 08544, USA}

\author{S. Torquato}
\affiliation{Department of Chemistry, Department of Physics, Princeton Institute for the Science and Technology of Materials, and Program in Applied and Computational Mathematics, Princeton University, New Jersey 08544, USA}


\date{\today}

%-------------------------------ABSTRACT-------------------------------------------

\begin{abstract}
    The concept of hyperuniformity has been a useful tool in the study of large-scale density fluctuations in systems ranging across the natural and mathematical sciences. One can rank a large class of hyperuniform systems by their ability to suppress long-range density fluctuations through the use of a hyperuniformity order metric $\bar{\Lambda}$. We apply this order metric to the Barlow packings, which are the infinitely degenerate densest packings of identical rigid spheres that are distinguished by their stacking geometries and include the commonly known fcc lattice and hcp crystal. The ``stealthy stacking" theorem implies that these packings are all stealthy hyperuniform, a strong type of hyperuniformity which involves the suppression of scattering up to a wavevector $K$. We describe the geometry of three classes of Barlow packings, two disordered classes and small-period packings. In addition, we compute a lower bound on $K$ for all Barlow packings. We compute $\bar{\Lambda}$ for the aforementioned three classes of Barlow packings and find that to a very good approximation, it is linear in the fraction of fcc-like clusters, taking values between those of least-ordered hcp and most-ordered fcc. This implies that the $\bar{\Lambda}$ of all Barlow packings is primarily controlled by the local cluster geometry. These results indicate the special nature of anisotropic stacking disorder, which provides impetus for future research on the development of anisotropic order metrics and hyperuniformity properties.
\end{abstract}

%-------------------------------ABSTRACT-------------------------------------------

\pacs{1}
%\keywords{}

\maketitle

%-------------------------------BODY-----------------------------------------------

\section{Introduction\label{intro}}

Continuing research into methods of characterizing density fluctuations has yielded many fundamental insights in science and mathematics \cite{density1,density2,density3,universe1,universe2}. Hyperuniformity has proven to be a useful framework for the investigation of the large-scale density fluctuations and structure of point patterns that arise in the physical, mathematical, and biological sciences \cite{hyp03, hyprev}. It generalizes a less visible property of long-range crystalline or quasicrystalline order, which is the suppression of density fluctuations at large length scales \cite{hyp03}. For a point pattern in $d$-dimensional Euclidean space $\mathbb{R}^d$, these fluctuations can be understood by computing the variance $\sigma^2_N(R)$ of the number of points inside a spherical window of radius $R$ as the window center is averaged over space \cite{hyp03}. While the variance of a Poisson or liquid-like point pattern grows as the window volume (i.e., $R^d$), that of a crystal grows as the surface area of the window (i.e., $R^{d-1}$) \cite{hyp03}. Torquato and Stillinger \cite{hyp03} generalized the difference in these cases by defining a \textit{hyperuniform} point pattern as one in which
\begin{equation}
    \lim_{R\to\infty} \frac{\sigma^2_N(R)}{v_1(R)} = 0,
\end{equation}
where $v_1(R)$ is the volume of a window of radius $R$. The hyperuniformity condition for a spherical window can be restated in Fourier space as the requirement that the structure factor $S(\mathbf{k})$, which is the elastic single scattering intensity of the point pattern, obey \cite{hyp03}
\begin{equation}
    \lim_{|\mathbf{k}|\to 0} S(\mathbf{k}) = 0.
\end{equation}
In addition to point patterns, hyperuniformity has been generalized to describe other important physical systems, including two-phase materials \cite{zquasi,hypgen} and scalar and vector fields \cite{hypgen,hypscalar}.

Importantly, hyperuniformity does not necessarily pose restrictions on the short-range order of the systems. Indeed, some of the most interesting hyperuniform systems have relatively low short-range order, such as the perfect glass, which is a hyperuniform, geometrically disordered, unique ground state of certain intrinsic two, three, and four-body interactions \cite{perfectglass}. Thus, hyperuniformity can often be described as a type of hidden order \cite{hidden}. Hyperuniformity arises in a variety of systems across multiple disciplines, including in the early density fluctuations of the universe \cite{universe1,universe2,universe3,universe4}, classical disordered ground states \cite{hidden,classical1,classical2,classical3,classical4,classical5,classical6,classical7,classical8,classical9,classical10}, maximally random jammed packings \cite{mrj1,mrj2,mrj3,mrj4,mrj5,mrj6}, models of plasmas \cite{plasma1, plasma2, plasma3, hyprev}, patterning of avian photoreceptor cells \cite{chicken}, quasicrystals \cite{zquasi, oquasi, lquasi}, and the spatial distribution of prime numbers \cite{primes1,primes2}. In addition, investigators have used the hyperuniformity concept as a tool to help design novel material properties such as isotropic photonic band gaps \cite{photonic1,photonic2,photonic3}, transparent dense disordered materials \cite{transparent}, desirable transport properties \cite{transport1,transport2}, and multifunctional materials \cite{multifunctional1,multifunctional2}. See Ref. \cite{hyprev} for a recent review of the basic theory and applications of hyperuniformity.

\begin{figure}
    \subfloat[][]{
        \includegraphics[width=0.6\linewidth]{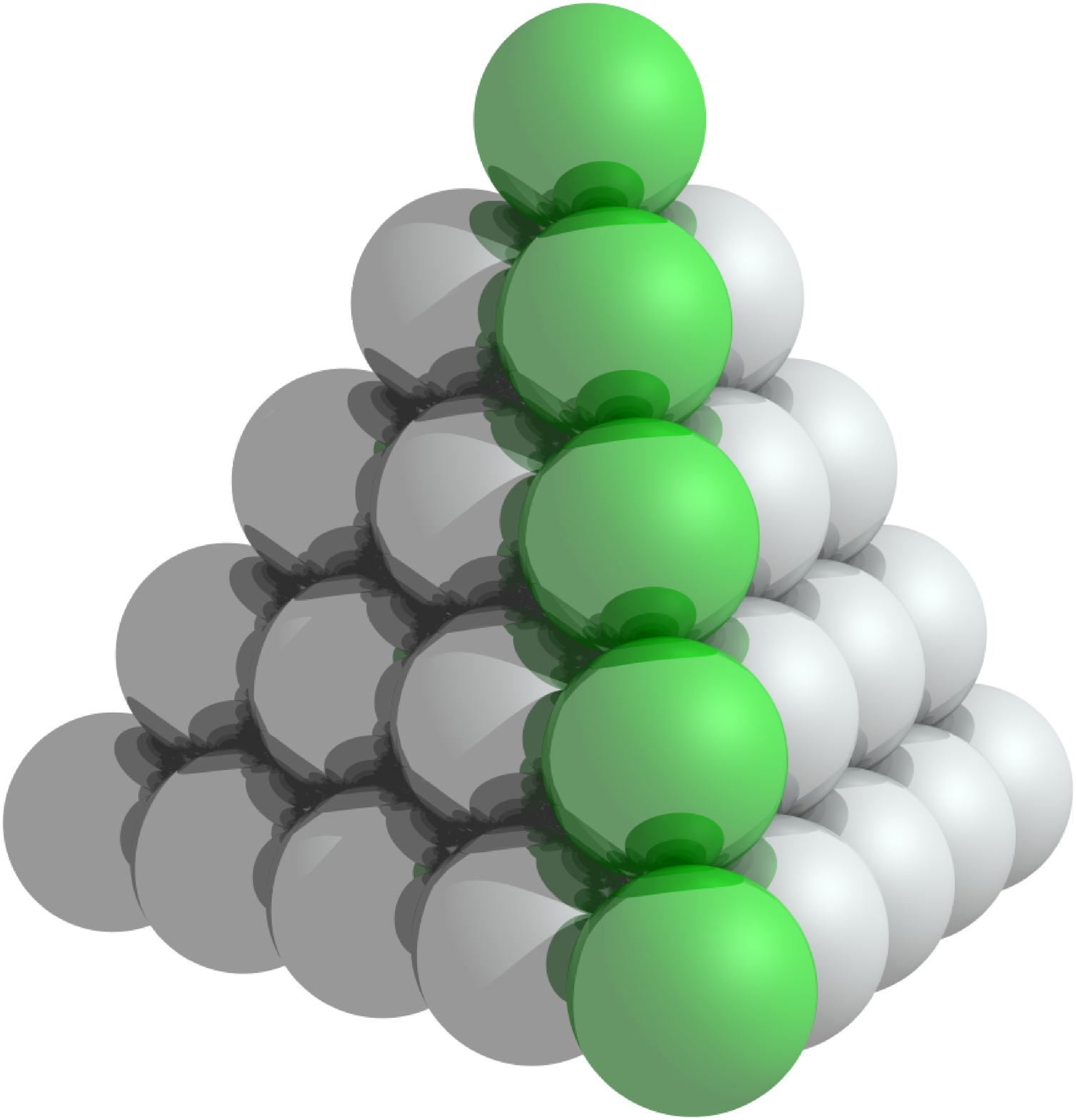}
    }\\
    \subfloat[][]{
        \includegraphics[width=0.6\linewidth]{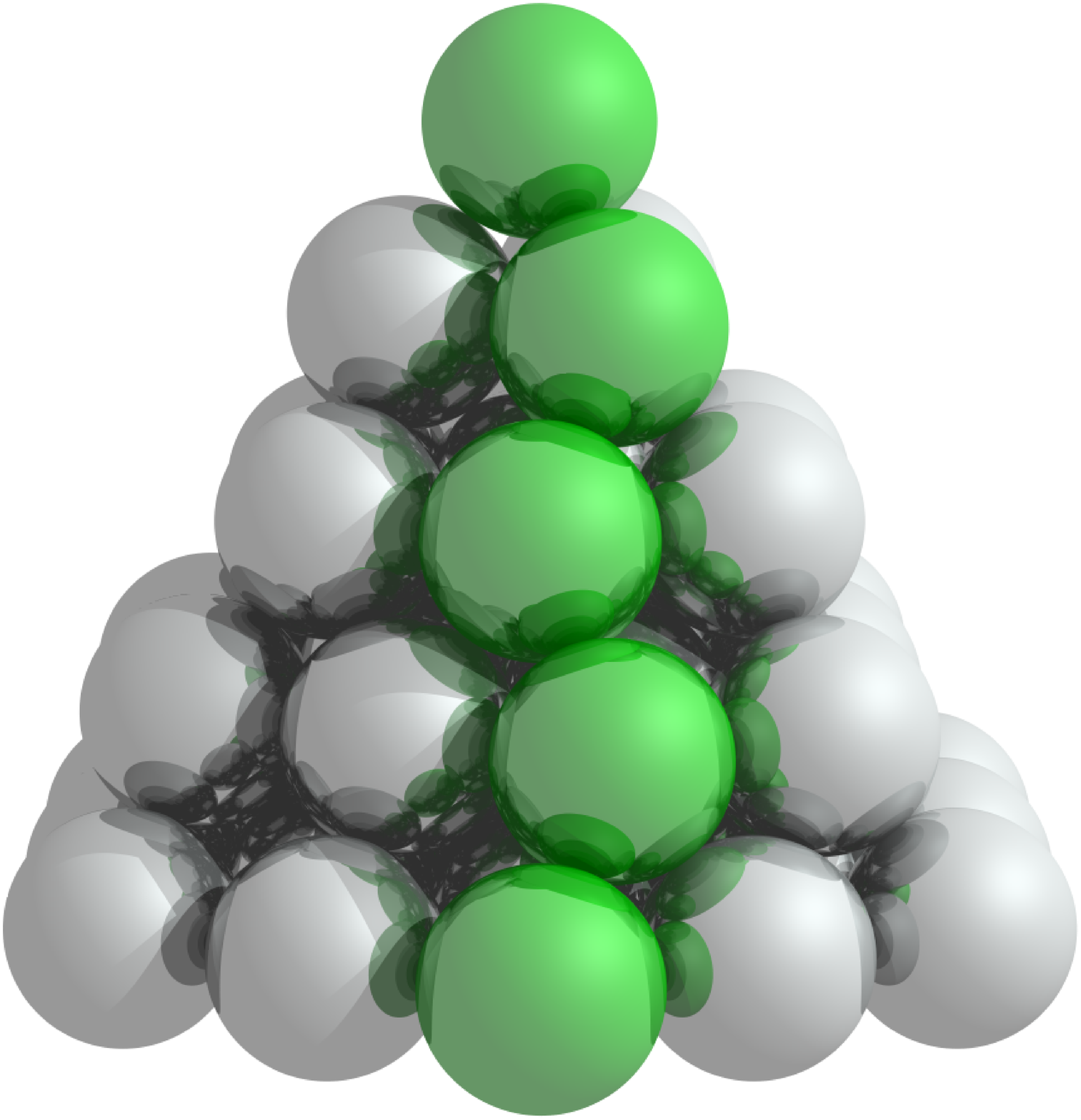}
    }
    \caption{A small fcc (a) and hcp (b) stacking. Notice how the green spheres in the fcc stacking rises in a straight line, while the hcp stacking has bends.\label{fccvhcp}}
\end{figure}

One interesting example of a class of hyperuniform systems are the close-packed rigid sphere packings in $\mathbb{R}^3$. These packings, known as the Barlow packings \cite{1barlow1883,2barlow1883,bestpackings,recentprogress} (or stacking variants in the physics literature), have the maximal packing fraction $\phi = \pi/\sqrt{18}$ by the Kepler conjecture, which was proved by Hales \cite{kepler}. The most commonly known examples of Barlow packings are the fcc lattice and the hcp crystal, shown in Fig. \ref{fccvhcp}. All the Barlow packings are strictly jammed \cite{jfcchcp,jammingrev}. A strictly jammed packing prohibits the simultaneous displacement of any subset of the spheres such that they lose contact with each other and the remaining spheres, as well as all volume-nonincreasing, uniform strains of the boundary of the packing \cite{jfcchcp,jammingrev}. In addition to being the maximally dense packings, when one removes spheres from them in specific ways, one can construct tunneled stacking variants that have a density of $\phi = \pi\sqrt{2}/9 = 0.49365\ldots$ \cite{tunneled}. These tunneled packings are believed to be the least dense strictly jammed packings \cite{tunneled}.

The Barlow packings have a variety of interesting physical properties. As one compresses an equilibrium hard-sphere system along the stable crystal branch, it must end in one of the strictly jammed Barlow packings \cite{rhm}. It is known by simulation that as the jammed state is approached, fcc wins over hcp \cite{eqsim1} by a relative free energy difference of order $10^{-3}$ \cite{eqsim1,eqsim2}. One can also consider a different type of ground state problem, where one wants to know the minimizer of a soft potential. In the space of lattices, fcc is a local minimizer of the inverse power law potential \cite{hyprev,epstein1,epstein2}
\begin{equation}
    \Phi(r) = \frac{1}{r^{s}},\qquad s > 3.\label{powerlawfcc}
\end{equation}

Remarkably, this last statement is closely related to hyperuniformity properties \cite{hyp03,hyprev}. To understand this point, consider the following large-$R$ asymptotic expansion of the variance for a certain class of hyperuniform point patterns introduced by Torquato and Stillinger \cite{hyp03}:
\begin{equation}
    \sigma^2_N(R) = \Lambda(R) \left(\frac{R}{D}\right)^{d-1} + o\left(\frac{R}{D}\right)^{d-1},\label{asymptotic}
\end{equation}
where $\Lambda(R)$ is generally a fluctuating function that must increase more slowly than linear in $R$, on average, and $D$ is a characteristic microscopic length scale. For a large class of systems, known as Class I hyperuniform systems, $\Lambda(R)$ is a bounded function that fluctuates about some average constant value \cite{zquasi,hyprev}. Class I hyperuniform systems include all perfect crystals \cite{hyp03}, many quasicrystals \cite{zquasi,oquasi,lquasi} and exotic disordered point patterns \cite{hyp03,classical1,classical2,classical3,hidden,perfectglass,universe4,plasma1,plasma2,plasma3}. One can rank such systems according to their ability to suppress large-scale density fluctuations using the \textit{hyperuniformity order metric} \cite{hyp03}
\begin{equation}
    \bar{\Lambda} = \lim_{L\to\infty} \frac{1}{L}\int_0^L \Lambda(R)\,dR.\label{average}
\end{equation}
Since $\bar{\Lambda}$ is the coefficient of fluctuation growth, we say that a system is more ordered with respect to large scale density fluctuations if it has a lower $\bar{\Lambda}$. To avoid problems inherent in comparing systems of different densities, we report all values of $\bar{\Lambda}$ for systems setting $D=1$ and rescaling to have number density $\rho = \sqrt{2},$ which is natural choice for describing the systems that are considered later in the article. The problem of minimizing $\bar{\Lambda}$ can be viewed as a ground-state optimization problem \cite{hyp03}, and the global minimum for $\bar{\Lambda}$ in three dimensions is currently believed to be achieved by the bcc lattice, with $\bar{\Lambda} = 1.01881$ \cite{hyp03,hyprev}. However, as we will review in Section \ref{prelim}, the dual lattice of the minimizer of $\bar{\Lambda}$ restricted to lattices is the minimizer of an inverse power-law potential in reciprocal space \cite{hyp03,hyprev}. Thus, the fact that fcc is the local minimizer of the potential given by Eq. (\ref{powerlawfcc}) is related to the fact that bcc is apparently the minimizer of $\bar{\Lambda}$ \cite{hyp03,hyprev}. It is known that $\bar{\Lambda}$ favors fcc over hcp, with Refs. \cite{hyp03,hyprev} giving $\bar{\Lambda}_{\text{fcc}}= 1.01944$ and $\bar{\Lambda}_{\text{hcp}} = 1.01957$ \cite{hyp03,hyprev}. The difference may seem small, but consider that this difference is only an order of magnitude smaller than the difference between the free energies separating fcc from hcp as one approaches jamming in equilibrium \cite{eqsim1, eqsim2}.

In this article, we describe how $\bar{\Lambda}$ ranks several larger classes of Barlow packings. In order to do this, we need to first describe how to differentiate the Barlow packings. The origin of the differences in the Barlow packings lies in their stacking geometries, which can be encoded as \textit{stacking codes} \cite{code1,code2,conwaysloane,stackex1,stackex2,guinier,wilson,bestpackings,recentprogress}. These will be described in more detail later in the article, but for now, notice how the fcc packing in Fig. \ref{fccvhcp} rises in a straight line, while the hcp packing has a zig-zag stacking. While fcc and hcp are the simplest stacking codes, nature is known to use more complicated periodic Barlow structures, such as in the crystal structures of various metals and metal compounds \cite{stackex1,stackex2}. A few examples have been listed in Table \ref{physicalex}. These stacking codes not only allow us to consider different periodic close-packed structures, but also allow us to introduce disorder in a very controlled manner by allowing the layers to be stacked probabilistically. There are also metallic and colloidal examples of stacking disordered Barlow packings \cite{wilson,exprandom1,exprandom2,exprandom3}, some of which are listed in Table \ref{physicalex}. While the space of Barlow packings is too large to be described exhaustively, we consider three specific classes with relatively simple parameterizations of the stacking geometry.

In order to compute $\bar{\Lambda}$, we must know that Barlow packings are Class I. This is guaranteed by a stronger condition \cite{hidden}, known as \textit{stealthy hyperuniformity}, which requires \cite{hidden}
\begin{equation}
    S(k) = 0,\qquad0\leq k<K,
\end{equation}
for some finite $K$. This condition holds trivially for any periodic structure with a finite basis, since the first feature in the spectrum is the first Bragg peak. When considering disordered systems, it is a general principle that any disordered system can be approximated by a large enough periodic system, even as the basis of the disordered system becomes infinitely large. However, the usual case for a non-stealthy disordered system is that when one tries to get progessively better approximations with larger periodic systems, the value of $K$ drops towards zero. Surprisingly, stealthy hyperuniformity holds rigorously for all Barlow packings, including the infinite stacking disordered ones with probabilistic codes. Furthermore, all Barlow packings share a common lower bound for $K$. This is a non-trivial statement that will be elaborated in Section \ref{stealthy}.

Confident in the validity of extending $\bar{\Lambda}$ computations to our three classes of Barlow packings, we can determine the ordering of Barlow packings with respect to the suppression of long-range fluctuations. While one might expect packings with a degree of disorder to possess larger long-range fluctuations, we show, counterintuitively, that these Barlow packings lie ``between" the fcc and hcp values, depending on the distribution of nearest-neighbor geometries, or cluster geometries. For the case of the Barlow packings, the rigid nature of the stacking geometry permits essentially two type of clusters, which are fcc-like and hcp-like \cite{wilson,guinier,recentprogress,bestpackings,conwaysloane}. More specifically, we show that $\bar{\Lambda}$ depends nearly linearly on the fraction of fcc-like clusters in the packing. Thus, $\bar{\Lambda}$ is relatively insensitive to long-distance developments in stacking complexity, and instead appears to rely more heavily on the geometry of local clusters. One can contrast this with the general development of hyperuniformity itself, which is an intrinsically long-range phenomenon, and cannot necessarily be predicted using local means.

\begin{table}
    \begin{tabular}{|c|c|c|}
        \hline
        Stacking Code & Systems & Refs.\\
        \hline
        ABAC & La, Pr, Nd, Am, Ce, HgBr$_2$, & \cite{stackex1,stackex2}\\
        & HgI$_2$, Ti$_2$S$_3$, Cd(OH)Cl & \\
        ABABCBCAC & Sm, Mo$_2$S$_3$, Li ($T = 4.2$ K) & \cite{stackex1, stackex2}\\
        ABABCBABAC & Ti$_4$S$_5$ & \cite{stackex2}\\
        ABCBCABABCAC & Fe$_3$S$_4$, Ti$_5$S$_8$ & \cite{stackex2}\\
        ``disordered variants'' & Co, Silica colloids & \cite{exprandom1,exprandom2, exprandom3,wilson}\\
        \hline
    \end{tabular}
    \caption{A list of examples of metals \cite{stackex1,stackex2,exprandom1,wilson}, metal compounds \cite{stackex2}, and colloidal \cite{exprandom2,exprandom3} systems that exhibit complex or disordered stacking codes. For the case of the metal compounds, the anions are arranged in a Barlow packing, with the cations occupying the holes of this packing \cite{stackex2}.\label{physicalex}}
\end{table}

This article is organized as follows. Section \ref{prelim} covers mathematical preliminaries such as stacking codes and the theory of hyperuniformity needed to understand the results of the article. Section \ref{class} introduces the three classes of Barlow packings we will use in our explicit computations: a stacking disordered system first described in Ref. \cite{wilson}, a class of ordered-disordered stacking mixtures, and the periodic codes with nine or fewer letters \cite{code1}. Section \ref{stealthy} applies the stealthy stacking theorem \cite{classical6} to derive a lower bound on $K$ for all Barlow packings, and comments on the realizability of these types of bounds. Section \ref{calculation} reports computations of $\bar{\Lambda}$ for the three classes of Barlow packings considered in this paper. In Section \ref{conclusion}, we summarize our findings and discuss their implications.


\section{Mathematical Preliminaries\label{prelim}}

In this section, we introduce a variety of mathematical concepts needed to understand the calculations in the latter part of the article.

\subsection{Packings, Lattices, and Crystals\label{packing}}

A \textit{sphere packing} is a collection of nonoverlapping spheres in $\mathbb{R}^d$. In this article, we consider only identical spheres of diameter $D$ in $\mathbb{R}^3$ and set $D=1$ without loss of generality. One important characteristic of a sphere packing is its packing fraction $\phi$, which is the fraction of space covered by the spheres. The maximal packing fraction $\phi_{\text{max}}$ in $\mathbb{R}^3$ is $\pi/\sqrt{18}$, as proven by Hales \cite{kepler}. Sphere packing models are useful for describing properties of dense many-body systems \cite{packingperspective} and probing certain mathematical problems \cite{conwaysloane} in which exclusion-volume effects play a dominant role. Examples include coding problems in signals theory \cite{conwaysloane}, the study of equilibrium phase transitions \cite{rhm,eqsim1,eqsim2}, and the study of jamming \cite{rhm,eqsim1,eqsim2,jfcchcp,jammingrev,mrj1,mrj2,mrj3,mrj4,mrj5,mrj6}.

A \textit{lattice} is a special case of a sphere packing that naturally introduces periodicity. In a $d$-dimensional lattice, the positions of all sphere centers are given by the integer sum of $d$ linearly independent vectors
\begin{equation}
    \mathbf{r} = \sum_{i=1}^d n_i \mathbf{v}_i, \qquad n_i\in\mathbb{Z}.\label{intsum}
\end{equation}
Common examples of lattices include the simple cubic lattice, the fcc lattice, and the body-centered cubic (bcc) lattice. Lattices play an essential role in number theory, where they are related to the study of quadratic forms \cite{conwaysloane}. For a review of the mathematical study of lattices and sphere packings, see Ref. \cite{conwaysloane}. Note that in the general physics literature, what we call a lattice is often known as a \textit{Bravais lattice}.

There is also the concept of a more general periodic point pattern, known as a \textit{crystal}. A crystal consists of the fundamental cell of a lattice $\Gamma$, into which a finite number $N\geq1$ of points are placed. This fixed configuration is then repeated over space by translating the fundamental cell by the lattice vectors of $\Gamma$. One way of representing a crystal formulaically is to add to the integer sum (\ref{intsum}) an extra vector $\mathbf{b}_n$
\begin{equation}
    \mathbf{r}_n = \sum_{i=1}^d n_i \mathbf{v}_i + \mathbf{b}_n,
\end{equation}
which represents the locations of the ``basis" particles. A crystal is then the union of the sets $\{\mathbf{r}_n\}$ for all $\mathbf{b}_n$, which are the spheres in the fundamental cell.

\subsection{Stacking Codes\label{codes}}

\begin{figure}
    \includegraphics[width=\linewidth]{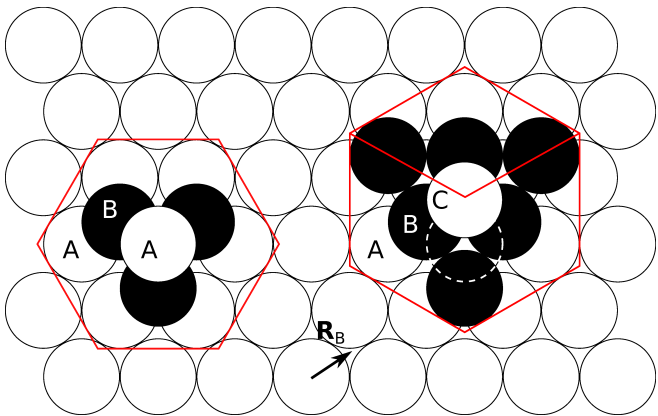}
    \caption{We can construct a Barlow packing by repeatedly stacking close packed layers. After putting down a layer of type $A$, there are two choices $B$ and $C$ for the next layer. The illustration is derived from public domain content hosted at \cite{wiki3}.\label{codefig}}
\end{figure}

The densest sphere packings in three dimensions are the infinitely degenerate Barlow packings. More precisely, the Barlow packings are the saturated (i.e., space does not exist to add any sphere without introducing overlap), strictly jammed densest packings \cite{jammingrev}. However, even within such constraints, there is still freedom in \textit{stacking}. This freedom is shown in Fig. \ref{codefig}. One begins by laying down a single two-dimensional close-packed layer, which we denote as type $A$. Then, one can lay a second two-dimensional close-packed layer in one of two sets of ``pockets," which we denote as type $B$ or $C$. The displacement of a type $B$ layer from the origin in the plane is $\mathbf{R}_B = (1/2,1/2\sqrt{3}),$ and is labeled in Fig. \ref{codefig} ($\mathbf{R}_C = -\mathbf{R}_B$). One can then continue this code to get a Barlow packing, subject to the constraint that there are no consecutive repeated letters. For example, the fcc lattice is given by the repeating code $ABC$, while the hcp crystal is given by the repeating code $AB$. These sequences can be infinite, leading to the conclusion that there are an uncountably infinite number of Barlow packings \cite{conwaysloane,recentprogress}. For the remainder of the article, we will also view any finite sequence as periodic, imposing an additional requirement that the last letter of the sequence is different from the first. We can also allow for the use of probabilistic codes, which is useful for defining ensemble averages over the Barlow packings.

Although we can enumerate all of the Barlow packings in this way, the codes generated are not unique, in the sense that multiple codes can refer to the same Barlow packing \cite{code1,code2,guinier}. We consider two packings to be the same if they are related by a simple translation or rotation. Previous investigators have developed an understanding of the codes that describe distinct Barlow packings \cite{code1,code2}, but for the purpose of this article it is sufficient to note that translation along the vector $\mathbf{R}_B$ allows us to begin all of our sequences with $A$.

The space of all Barlow packings is large, and we will not be able to exhaustively discuss it in this article. As such, we will focus on three specific subclasses of Barlow packings: an infinite packing composed of uncorrelated fcc and hcp clusters, a class of order-disorder mixture packings, and the periodic Barlow packings of code length up to nine. The geometry of these specific packings is discussed in Section \ref{class}.

\subsection{Theta Series and Lattice Sums\label{theta}}

A fundamental geometric quantity of interest in this article is the ensemble-averaged pair correlation function in the infinite system limit:
\begin{equation}
    g_2(\mathbf{r}) = \frac{1}{\rho}\sum_{i\neq 0} P_i \delta(\mathbf{r} - \mathbf{r}_i),\label{g2}
\end{equation}
where $P_i$ is the probability that we find a particle at $\mathbf{r}_i$ given that the origin is taken as random particle in the packing, $\rho$ is the number density, and the sum runs over all possible sites except for the origin. While we do consider unique periodic point configurations in this article, one can still use an ensemble average formulation of the pair statistics by defining the ensemble average to be taken over the particles in the fundamental cell with equal weight.

As a practical matter, descriptors derived from this quantity can be computed by lattice sums. A lattice sum is simply the process of evaluating a sum by explicitly enumerating over points in the lattice and adding the contributions to the desired function. Most often, we do this numerically. There is, however, a closely related computational tool from the mathematical theory of lattices, known as a \textit{theta series}. The theta series of a lattice $\Gamma$ can be defined as \cite{conwaysloane}
\begin{equation}
    \theta_\Gamma(q) = \sum_{j=0}^\infty Z_j q^{r_j^2},\label{thetadef} 
\end{equation}
where $Z_{j}$ is the coordination number of shell $j$ around the particle at the origin. We also need the analogous concept of a theta series around an arbitrary point with respect to the lattice \cite{conwaysloane}, which just involves computing $Z_j$ and $r_j$ from the perspective of an arbitrary point.

One can easily generalize the theta series of a lattice in the following manner. If we take an angular average of $g_2$, we can write it in the form \cite{hyprev}
\begin{equation}
    g_2(r) = \frac{1}{4\pi\rho} \sum_{j=1}^\infty \frac{Z_j}{r_j^2} \delta(r - r_j),\label{isotropicg2}
\end{equation}
where $j$ runs over all possible coordination shells, beginning at the nearest neighbor, and $Z_j$ is the expected coordination number in that shell \cite{hyprev}. We then can associate the $Z_j$ of Eq. (\ref{isotropicg2}) with that of Eq. (\ref{thetadef}) to get the \textit{average theta series} of a packing (setting $Z_0 = 1$) \cite{conwaysloane}. We can also define the concept of a theta series for a periodic packing around an arbitrary point in the fundamental cell, which just involves computing $Z_j$ from the perspective of an arbitrary point without averaging. 

Many interesting identities have been derived for series of this type \cite{conwaysloane,sloanediamond}. As a concrete example of one way in which these functions can be manipulated, consider the problem of obtaining the theta series of a lattice $\Gamma'$ dilated from a lattice $\Gamma$ by a factor $\lambda$, given the original theta series $\theta_\Gamma$. The new theta series is then
\begin{equation}
    \theta_{\Gamma'}(q) = \theta_{\Gamma}(q^{\lambda^2}).
\end{equation}
We will use similar observations to derive an expression for the theta series of certain classes of Barlow packings in Section \ref{class}.

\subsection{Hyperuniformity Order Metric\label{metric}}

As stated in Section \ref{intro}, it is possible to define a hyperuniformity order metric $\bar{\Lambda}$ for Class I hyperuniform systems in terms of an asymptotic expansion of the number variance (\ref{asymptotic}). This hyperuniformity order metric rank orders systems by their ability to suppress long-range fluctuations, and can be thought of as the average rate of long-range fluctuation growth for these systems. In this section, we present the derivation of an explicit formula for the constant  $\bar{\Lambda}$ in terms of the ensemble-averaged pair correlation function $g_2({\bf r})$, which can be computed through methods presented in the previous section.

We begin with the definition of the number variance in terms of the number of points $N(R)$ contained inside a hyperspherical window of radius $R$:
\begin{equation}
    \sigma^2_N(R) = \langle N^2(R)\rangle - \langle N(R)\rangle^2.
\end{equation}
For a statistically homogeneous system \footnote{Strictly speaking, the Barlow packings are statistically inhomogeneous. However, this formula can also be used for the case of a crystal, as long as one interprets the ensemble average to be taken over the different points in the fundamental cell. For the classes of disordered packings we consider, the lattice translational invariance of the considered stacking probability functions will guarantee the validity of these formulas.}, one can show that these ensemble averages can be written as \cite{hyp03,hyprev}:
\begin{equation}
    \sigma^2_N(R) = \rho v_1(R) \left[1 + \rho \int h({\bf r}) \alpha_2({\bf r};R)\,d{\bf r}\right],\label{enssigma}
\end{equation}
where we have introduced the total correlation function $h({\bf r}) = g_2({\bf r}) - 1$ and the scaled intersection volume $\alpha_2({\bf r};R)$, which is the intersection volume of two spheres of radius $R$ separated by a displacement vector $\bf{r}$ normalized by $v_1(R)$. To obtain an asymptotic expansion of this expression, Torquato and Stillinger \cite{scaledintersection} showed that the scaled intersection volume for $r \leq 2R$ can be written in a series:
\begin{widetext}
\begin{equation}
    \begin{aligned}
        \alpha_2(r;R) &= 1 - c(d)x + c(d)\sum_{n=2}^\infty (-1)^n \frac{(d-1)(d-3)\cdots(d-2n+3)}{(2n-1)[2\cdot4\cdot6\cdots(2n-2)]} x^{2n-1},\label{series}
    \end{aligned}
\end{equation}
\end{widetext}
where $x = r/2R$ and
\begin{equation}
    c(d) = \frac{2\Gamma(1 + d/2)}{\pi^{1/2}\Gamma[(d+1)/2]}.
\end{equation}
This series is truncated for odd dimensions, but is an infinite series for even dimensions \cite{hyprev}. Insertion of Eq. (\ref{series}) into Eq. (\ref{enssigma}) gives the expansion in Eq. (\ref{asymptotic}), with \cite{hyprev}
\begin{equation}
    \Lambda(R) = -\frac{2^d\phi^2 c(d)}{2 D v_1(D/2)}\int_{r\leq 2R} h({\bf r}) |{\bf r}|\,d{\bf r}.\label{nonaveraged}
\end{equation}
Finally, assuming Class I hyperuniformity and the existence of the limit of $\Lambda(R)$ as $R\to\infty$, one can take the average in Eq. (\ref{average}) simply by taking a limit to get \cite{hyprev}
\begin{equation}
    \bar{\Lambda} = -\frac{2^d\phi^2 c(d)}{2 D v_1(D/2)}\int h({\bf r}) |{\bf r}|\,d{\bf r}.
\end{equation}
However, if we insert the expansion (\ref{isotropicg2}) into this expression, we do not obtain individually convergent terms. One can produce a convergent expression by introducing a Gaussian factor and taking the limit as the Gaussian approaches a uniform function, and arrive at the sum \cite{hyp03,hyprev}
\begin{equation}
    \begin{aligned}
    \bar{\Lambda} &= \lim_{\beta\to 0^+}\frac{2^{d-1}\phi d}{D \Gamma(1/2)}\\
        &\times\left[\frac{\phi \pi^{d/2}}{v_1(D/2)\beta^{\frac{d+1}{2}}} - \frac{\Gamma(d/2)}{\Gamma(\frac{d+1}{2})}\sum_{j=1}^\infty Z_j r_j e^{-\beta r_j^2}\right].\label{extrapolation}
    \end{aligned}
\end{equation}
This equation can be used to estimate $\bar{\Lambda}$ by extrapolating computations done with finite $\beta$ to $\beta = 0$. One can also use an expression obtained by substituting Eq. (\ref{nonaveraged}) into Eq. (\ref{average}) without the assumption that $\Lambda(R)$ has a limit as $R\to\infty$. Reference \cite{windowshape} discusses one approach to understanding the difference between these two methods. We will use Eq. (\ref{extrapolation}) exclusively in this article.

It is worth noting that one can instead do computations for $\bar{\Lambda}$ in Fourier space using the structure factor
\begin{equation}
    S({\bf k}) = 1 + \rho \tilde{h}({\bf k}),\label{structurefactordef}
\end{equation}
where $\tilde{h}({\bf k})$ is the Fourier transform of $h({\bf r})$. One can use Parseval's theorem to rewrite Eq. (\ref{enssigma}) \cite{hyprev}:
\begin{equation}
    \sigma^2_{N}(R) = \rho v_1(R) \left[\frac{1}{(2\pi)^d} \int S({\bf k}) \tilde{\alpha}_2({\bf k};R)\,d{\bf k}\right],
\end{equation}
where $\tilde{\alpha}_2({\bf k};R)$ is the Fourier transform of the scaled intersection volume. Taking an asymptotic expansion of this function and averaging, we obtain \cite{hyprev}
\begin{equation}
    \bar{\Lambda} = \frac{\rho \Gamma(1 + d/2) v_1(1) D^{2d}}{\pi^{1 + d/2}}\int \frac{S({\bf k})}{(kD)^{d+1}}\,d{\bf k}.
\end{equation}
This formula can be interpreted as an energy evaluation of the dual configuration in a power law potential, provided that $S({\bf k})$ can itself be interpreted as a direct space pair correlation function, i.e., $S(\bf{k})$ has a constant intensity where it is supported. While this realizability condition does not hold generally, it is true for certain configurations including lattices, some crystals \cite{dual1,dual2}.

\section{Three Classes of Barlow Packings\label{class}}

In this section, we begin by introducing a local cluster statistic, and then use this cluster statistic to define a class of disordered Barlow packings previously used in Refs. \cite{wilson,guinier}. This class of packings essentially consists of uncorrelated fcc and hcp layers. We also introduce a class of order-disorder mixture packings consisting of random choices of stacking between fixed $A$ layers. We then discuss some of the properties of small code length periodic Barlow packings. We will take all packings to have particle diameter $D = 1$.

\subsection{Local Clusters\label{local}}

\begin{figure}
    \subfloat[][]{
        \includegraphics[width=0.5\linewidth]{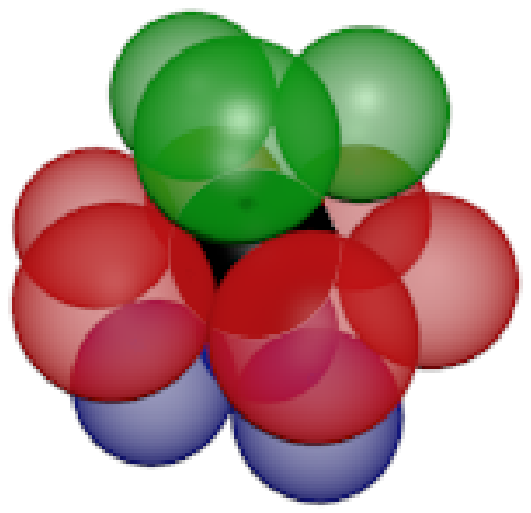}
    }%
    \subfloat[][]{
        \includegraphics[width=0.5\linewidth]{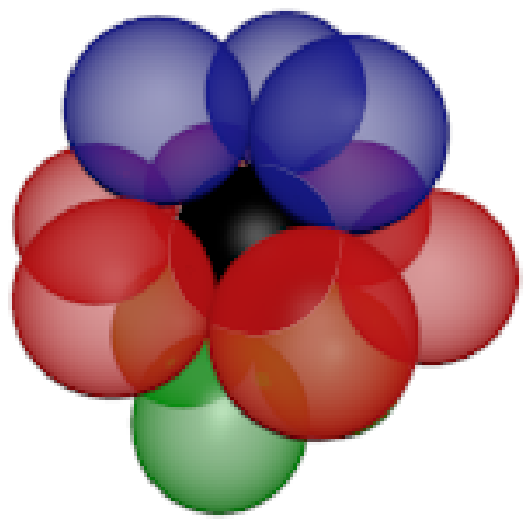}
    }\\
     \subfloat[][]{
        \includegraphics[width=0.5\linewidth]{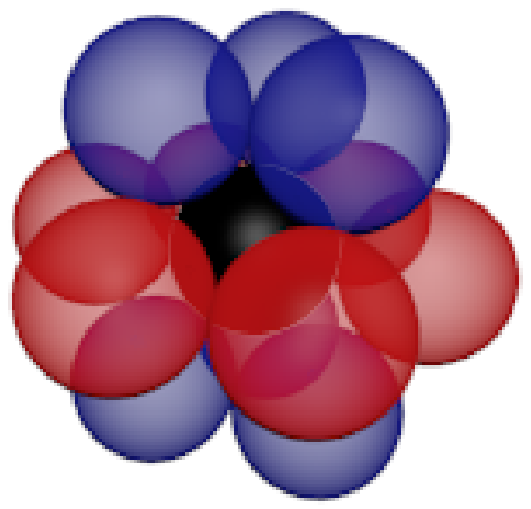}
    }%
    \subfloat[][]{
        \includegraphics[width=0.5\linewidth]{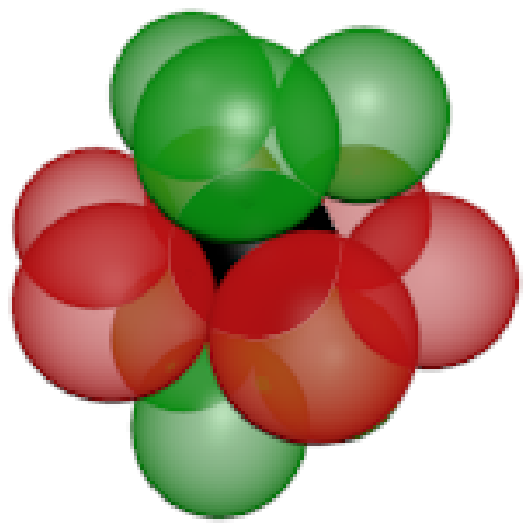}
    }
    \caption{The four types of local clusters. Panel (a) is fcc, panel (b) is reverse fcc, panel (c) is hcp, and panel (d) is reverse hcp.\label{clusterfig}}
\end{figure}

Figure \ref{clusterfig} depicts the four cluster types that can arise in an arbitrary Barlow packing. They are the ABC (fcc), CBA (reverse fcc), ABA (hcp), and ACA (reverse hcp) clusters. The ABC and CBA clusters are related by a mirror plane, while the ABA and ACA clusters are related by an inversion. Since we can swap all forward and reverse clusters in a packing by a rotation, all physically relevant properties may at most rotate under such a swap. All of the quantities we will consider in this article have the even stronger property of complete insensitivity to the difference between forward and reverse clusters, so we will ignore the difference and denote the fraction of fcc-type clusters as $\alpha$. In this way, the fcc and hcp packings can be seen as the two extremes of the Barlow packings, with respect to local cluster statistics. However, this fraction cannot completely parameterize the Barlow packings. As an example, consider that the codes ABABAC and ABCBCACAB both have $\alpha = 1/3$.

\subsection{Uncorrelated FCC and HCP Clusters\label{uncorrelated}}

We can use the local cluster statistic $\alpha$ of the previous section to define a simple class of infinite Barlow packings, which are composed of uncorrelated fcc and hcp-like clusters. This construction first appeared in Ref. \cite{wilson}, where it was used to help model the structure of cobalt.

The strategy is to take an ensemble approach to define this packing, and so we seek to compute $P_{A}(m, \alpha)$, which is the probability that a layer $m$ spaces away from our arbitrary starting layer is type $A$ given a fcc fraction of $\alpha$. We can always choose the starting layer as type $A$, so $P_{A}(0, \alpha) = 1$. Then, the next layer will always be $B$ or $C$, so $P_A(1, \alpha) = 0$. Afterwards, the layer probabilities will follow the recursion relation \cite{wilson,guinier}
\begin{equation}
    \begin{aligned}
        P_{A}(m,\alpha) &= \alpha [1- P_A(m-1,\alpha) - P_A(m-2,\alpha)]\\
        &+ (1-\alpha)P_A(m-2,\alpha),
    \end{aligned}
\end{equation}
since if the cluster is fcc-like, the layer will be type $A$ if and only if both layers before are not $A$, and if the cluster is hcp-like, the layer will be type $A$ if and only if the layer 2 spaces before is $A$. Since each cluster is independent of the one before it, we call this the uncorrelated cluster class of infinite Barlow packings.

\begin{figure}
    \includegraphics[width=\linewidth]{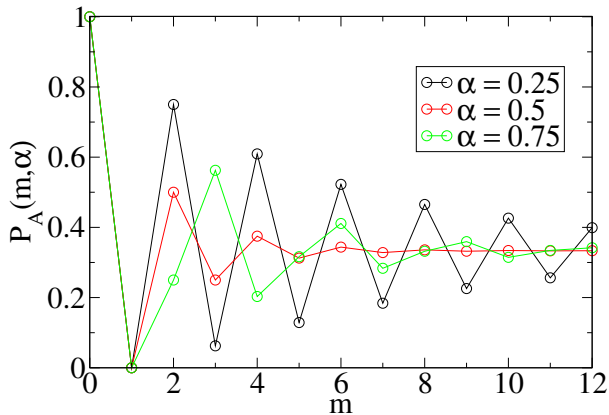}
    \caption{This plots the function $P_A(m, \alpha)$ for $\alpha = 0.25, 0.5,$ and $0.75$. Notice how the correlations are smaller for $\alpha = 0.5$, reflecting a more disordered stacking.\label{probfig}}
\end{figure}

To solve the above recursion relation, one uses an ansatz $P_A(m,\alpha) = A x^{m} + By^{m} + C$ \cite{wilson,guinier,milne} and the observation that one can extend the solution to negative $m$ on physical grounds by requiring symmetry about $m=0$ \cite{wilson,guinier}. The solution is \cite{wilson,milne}
\begin{widetext}
\begin{equation}
    \begin{aligned}
        P_{A}(m,\alpha) &= \frac{1}{3} + \left( \frac{1-\alpha + \sqrt{4-8\alpha + \alpha^2}}{3\sqrt{4-8\alpha+\alpha^2}}\right)\left[-\frac{1}{2}\left(\alpha+\sqrt{4-8\alpha+\alpha^2}\right)\right]^{|m|}\\
        &+ \left(\frac{-1+\alpha + \sqrt{4-8\alpha + \alpha^2}}{3\sqrt{4-8\alpha+\alpha^2}}\right) \left[\frac{1}{2}\left(-\alpha+\sqrt{4-8\alpha+\alpha^2}\right)\right]^{|m|}.
    \end{aligned}
\end{equation}
\end{widetext}
Note that a limiting procedure needs to be used at some values of $\alpha$ and $m$, but this is easy to overcome in practice, as it occurs at relatively isolated points. This probability function can be used to compute an average theta series or carry out a lattice sum, which is sufficient to evaluate the descriptors used in this article. For visual reference, this function has been plotted for three values of $\alpha$ in Fig. \ref{probfig}.

This construction is completely symmetric under the exchange of layers $B$ and $C$. In general, packings such as the fcc packing rotate under exchange of $B$ and $C$, but this poses no problems for the types of geometric descriptors we seek to compute. The $\alpha = 0$ limit is the average of the $AB$ and $AC$ hcp packing, while the $\alpha = 1$ limit is the average of the forward and reverse fcc packing. These are the exchange invariant combinations, but they can still be thought of as the pure fcc and hcp packings for the purposes of this article. The $\alpha = 1/2$ limit is also noteworthy, as we consider this the ``most disordered" Barlow packing, giving a precise definition to the notion introduced in Ref. \cite{jammingrev} and making contact with the discussion in Ref. \cite{guinier}.

One can build the theta series of the uncorrelated cluster class from known theta series for two-dimensional lattices through addition and shift operations, a general strategy taken from Refs. \cite{conwaysloane,sloanediamond}. If one knows that the desired $g_2$ of packing $\Gamma$ is composed of the weighted sum of the local density of (possibly translated) lattices $\{\Gamma_{i}\}$, then one has that \cite{conwaysloane,sloanediamond}
\begin{equation}
    \theta_{\Gamma}(q) = \sum_i w_i \theta_{\Gamma_i}(q).
\end{equation}
If one translates a two-dimensional lattice $\Gamma$ in a direction orthogonal to the extent of the lattice (we take this as the $z$ direction), the theta series of the translated lattice $\Gamma'$ around the same origin is
\begin{equation}
    \theta_{\Gamma'}(q) = q^{z^2} \theta_{\Gamma}(q).
\end{equation}
These two observations can be combined to give a formula for the theta series ($\theta_\alpha$) of the uncorrelated cluster class in terms of $P_A(m,\alpha)$, the theta series for a triangular lattice viewed from a lattice point ($\theta_\text{point}$) and central hole ($\theta_\text{hole}$)\footnote{One can find the formulas for $\theta_{\text{point}}$ and $\theta_{\text{hole}}$ in terms of more typical elliptic theta functions in Refs. \cite{conwaysloane,sloanediamond}.}:
\begin{equation}
    \begin{aligned}
    \theta_{\alpha}(q) &= \sum_{m=-\infty}^\infty q^{2m^2/3}\\
        &\times\left[P_A(m,\alpha)\theta_{\text{point}}(q) + (1-P_A(m,\alpha))\theta_{\text{hole}}(q)\right].
    \end{aligned}
\end{equation}
There is a nontrivial leftover explicit sum in the theta series, which we could not express in terms of relatively simple theta functions. It is possible to express it as a function known in the mathematical literature as a partial theta series \cite{partialtheta}, but doing so seems to offer no computational benefit.

As a concrete example of this theta series, the first few terms for $\alpha = 1/2$ are 
\begin{equation}
    \theta_{1/2}(q) =1 + 12q + 6q^2 + q^{8/3} + 21q^3 + \cdots,
\end{equation}
which one can compare with the known theta series for the hcp crystal ($\alpha = 0$) \cite{conwaysloane,tunneled}
\begin{equation}
    \theta_0(q) =1 +  12q + 6q^2 + 2q^{8/3} + 18q^3 + \cdots,
\end{equation}
and the fcc lattice ($\alpha = 1$) \cite{conwaysloane,tunneled}
\begin{equation}
    \theta_1(q) = 1 + 12q + 6q^2 + 24q^3 + \cdots.
\end{equation}
This theta series has been used to carry out the numerical uncorrelated cluster family computations described in Section \ref{calculation}. The general strategy is to use \textit{Mathematica} to obtain partially subsituted symbolic expressions for coefficients up to $q^{5625}$, and then to numerically evaluate those expressions with Mathematica.

\subsection{Another Disordered Class}

We also consider another type of disordered packing, which introduces long-range stacking order while still keeping a degree of stacking disorder. This packing consists of an infinite array of $A$ layers, placed one space apart. In between, we place $B$ and $C$ layers randomly, with equal weight. We will call this the \textit{hcp order-disorder mixture packing}. This packing has $\alpha = 1/4$, and a theta series
\begin{equation}
    \theta_{\text{mhcp}}(q) = \frac{3}{4}\theta_{0}(q) + \frac{1}{4}\theta_C(q),
\label{odmixture}\end{equation}
where $\theta_0$ is the hcp theta series and $\theta_C$ is the theta series of a single layer of $C$ in a packing consisting of otherwise $A$s and $B$s, centered on a layer $C$ particle. The first few terms of this theta series are
\begin{equation}
    \theta_{\text{mhcp}}(q) = 1 + 12q + 6q^2 + \frac{3}{2}q^{8/3} + \frac{39}{2}q^{3}+\cdots.
\end{equation}
Note that this packing is ``hcp-like" in the sense that the $A$ layers are always spaced one apart from each other. We can also define an ``fcc-like" equivalent, where there are an infinite array of $A$ layers, with two spaces in between. These spaces can by randomly filled with $BC$ and $CB$ codes with equal weight. For this packing, we find $\alpha = 5/6$ and
\begin{equation}
    \theta_{\text{mfcc}}(q) = \frac{1}{2}\theta_{1}(q) + \frac{1}{3}\theta_C(q) + \frac{1}{6}\theta_{\text{dhex}}(q),
\end{equation}
where $\theta_{\text{dhex}}(q)$ is the theta series corresponding to a sequence $\ldots AAABABAAA\ldots$ (layers of $A$ continuing after ellipses) centered on the $A$ layer between the $B$ layers. The first few terms of this series are
\begin{equation}
    \theta_{\text{mfcc}}(q) = 1 + 12q + 6q^2 + \frac{1}{3}q^{8/3} + 20 q^3 + \cdots,
\end{equation}
We can also generalize these two packings to use unevenly weighted distributions for the random layers.

\subsection{Periodic Packings\label{periodic}}

\begin{table}
    \begin{tabular}{|c|c|c|}
        \hline
        Stacking Code  & $\alpha$ \\
        \hline
        $AB$ & 0\\
        $ABC$ & 1\\
        $ABAC$ & 1/2\\
        $ABABC$ & 3/5\\
        $ABABAC$ & 1/3\\
        $ABACBC$ & 2/3\\
        $ABABABC$ & 3/7\\
        $ABABCAC$ & 3/7\\
        $ABACABC$ & 5/7\\
        $ABABABAC$ & 1/4\\
        $ABABACAC$ & 1/4\\
    \hline
    \end{tabular}
    \begin{tabular}{|c|c|c|}
        \hline
        Stacking Code & $\alpha$ \\
        \hline
        $ABABACBC$ & 1/2\\
        $ABABCABC$ & 3/4\\
        $ABABCBAC$ & 1/2\\
        $ABACBABC$ & 3/4\\
        $ABABABABC$ & 1/3\\
        $ABABABCAC$ & 1/3\\
        $ABABACABC$ & 5/9\\
        $ABABCABAC$ & 5/9\\
        $ABABCACBC$ & 5/9\\
        $ABABCBCAC$ & 1/3\\
        $ABACBACBC$ & 7/9\\
        \hline
    \end{tabular}
    \caption{This table gives the unique periodic Barlow codes up to 9 letters long, along with their corresponding values of $\alpha$, the fcc cluster fraction. These sequences were compiled by extracting them from the list in Ref. \cite{code1} and removing duplicates, such as $ABAB$.\label{periodiccodes}}
\end{table}

Finally, we will work with the set of all periodic Barlow packings with codes up to 9 letters long. Since stacking codes are not unique, we will need to use a single representative per Barlow packing. A list of such representatives can be found in Ref. \cite{code1}, and we have reproduced the relevant part of the list in Table \ref{periodiccodes}. Note that Table \ref{periodiccodes} contains some of the periodic codes listed in Table \ref{physicalex}, the list of natural examples of periodic Barlow packing. We work with periodic Barlow packings in order to understand how $\bar{\Lambda}$ is affected by gradually increasing complexity, which is complementary to the types of information learned by considering the uncorrelated cluster and order-disorder mixture packings.

\section{Stealthy Hyperuniformity of Barlow Packings\label{stealthy}}

Reference \cite{classical6} proves a result known as the ``stealthy stacking" theorem and uses it to imply that all Barlow packings are stealthy. In this section, we briefly derive the resulting common lower bound on $K$ among all Barlow packings, and give further commentary on the realizability of these lower bounds. Consider a $(d_P + d_Q)$-dimensional Euclidean space $E$. The stealthy stacking theorem states that if one has a stealthy point pattern $P$ (up to $K_P$) in a $d_P$-dimensional subspace $E_P$, and a set of stealthy point patterns $\{Q(\mathbf{p}): \mathbf{p}\in P\}$ (with smallest stealthy limit $K_Q$) of common density $\rho_Q$ in the $d_Q$-dimensional orthogonal complement $E_Q$, then the point pattern in $E$ given by
\begin{equation}
    \mathbf{e} = \mathbf{p} + \mathbf{q},\qquad \mathbf{p}\in P,\, \mathbf{q}\in Q(\mathbf{p}),
\end{equation}
is stealthy, and the smaller of $K_P$ and $K_Q$ is a lower bound on $K$. This lower bound is not necessarily realized. This can be seen by considering the hexagonal lattice with a nearest neighbor distance of one (unit spacing), which is stealthy up to $K = 4\pi/\sqrt{3}$, while the lower bound provided by the stealthy stacking theorem by considering the integer lattice of spacing $\sqrt{2/3}$ as $P$ and the displaced integer lattices of unit spacing as $\{Q\}$ is $2\pi$. However, the lower bound is realized if $K_P$ is the lower bound.

For the case of the Barlow packings of unit spacing, one can consider the integer lattice of spacing $\sqrt{2/3}$ in the stacking direction to be $P$. This lattice is stealthy up to $K_P = \sqrt{6}\pi$. Then, there are three point patterns in the set $\{Q\}$, which are all displaced hexagonal lattices of unit spacing. These point patterns are all stealthy up to $K_Q = 4\pi/\sqrt{3}.$ Thus, all Barlow packings are stealthy, with $4\pi/\sqrt{3}$ as a lower bound on $K$.

This result can also be derived by directly computing $S(\mathbf{k})$ for an arbitrary Barlow packing. The interested reader can refer to Refs. \cite{wilson, guinier} for the basic strategy.

\section{Hyperuniformity Order Metric\label{calculation}}

\begin{figure}
    \subfloat[][]{
        \includegraphics[width=0.5\textwidth]{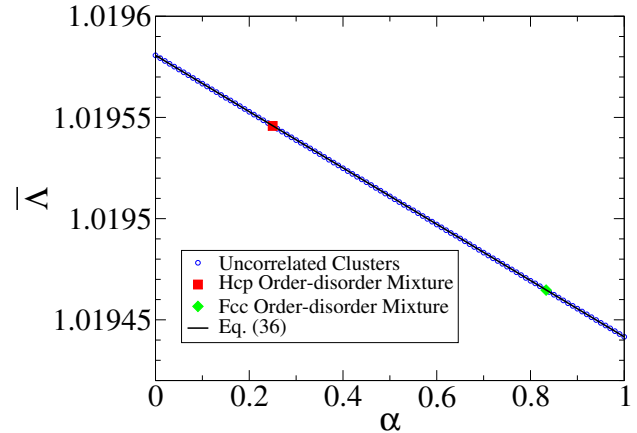}
    }\\
    \subfloat[][]{
        \includegraphics[width=0.5\textwidth]{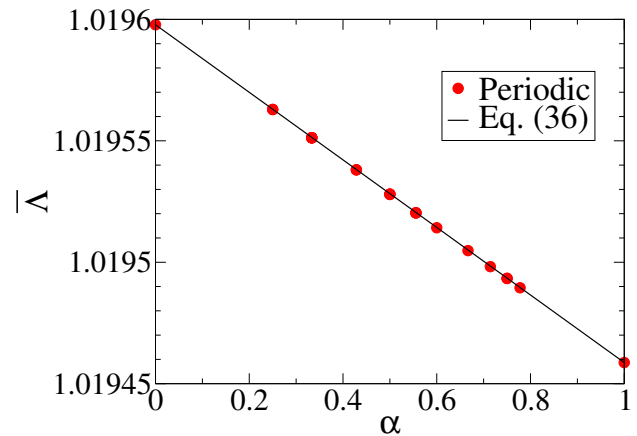}
    }
    \caption{These figure plot the model (\ref{linearmodel}) and the computed values of $\bar{\Lambda}$. Panel (a) gives the stacking disordered data, with the uncorrelated cluster model in blue, the hcp order-disorder mixture in red, and the fcc order-disorder mixture in green. Panel (b) gives the periodic Barlow data. Note that there are actually 22 data points in panel (b), but the differences between some points are indiscernibly small on the scale of the plot. The values of $\bar{\Lambda}$ needed for the model (\ref{linearmodel}) are determined by the endpoints of each plot.\label{data}}
\end{figure}

We present the results of our calculations $\bar{\Lambda}$ for the three classes of Barlow packings introduced in Section \ref{class}. The strategy for computing $\bar{\Lambda}$ for packings with stacking disorder is to compute the coefficients of the theta series up to $q^{5625}$ through the use of \textit{Mathematica}. Once this is done, we can estimate $\bar{\Lambda}$ by using Eq. (\ref{extrapolation}) with non-zero $\beta$ and linearly extrapolating to $\beta = 0$. We use a range of $\beta$ from 0.01 to 0.05, since the sum in Eq. (\ref{extrapolation}) converges before using all of the computed coefficients for those values of $\beta$. The typical set of data contains a slight degree of curvature such that a linear extrapolation underestimates $\bar{\Lambda}$ on the order of $10^{-5}$; see the Appendix for analyses of the fit and residual errors. For the order-disorder mixture, we computed the contributions from $\theta_{\text{hcp}}$ and $\theta_C$ separately, and combined them with the appropriate weight given in Eq. (\ref{odmixture}).

In contrast, we compute $\bar{\Lambda}$ for the periodic packings with codes up to 9 letters long (Table \ref{periodiccodes}) through a lattice sum procedure. Like the calculation for the stacking disordered packings, we use Eq. (\ref{extrapolation}) and linearly extrapolate to $\beta = 0$. Since our implementatation of this method of calculation is faster than our implementation using a theta series, we can use $\beta = 0.0005$-$0.0025$. This also requires the use of an arbitrary precision numerical library, as we found that floating point error with conventional 64 bit floating point numbers limited the precision. We use rug, a Rust wrapper around MPFR, with a mantissa of 128 bits. The fits are qualitatively similar to the stacking disordered packing fits, but we are only underestimating on the order of $10^{-7}$. Thus, we have improved on the computations of Ref. \cite{hyp03}, and find that $\bar{\Lambda}_{\text{fcc}} = 1.0194587$ and $\bar{\Lambda}_{\text{hcp}} = 1.0195978.$

We plot the estimated value of $\bar{\Lambda}$ against the fcc cluster fraction $\alpha$ for both the stacking disordered and periodic packings in Fig. \ref{data}, which is well modeled by the following simple linear weighted average:
\begin{equation}
    \bar{\Lambda}_{\text{lin}}(\alpha) = \alpha\bar{\Lambda}_{\text{fcc}} + (1-\alpha) \bar{\Lambda}_{\text{hcp}}.\label{linearmodel}
\end{equation}
Indeed these deviations are on the order of 1000 times smaller than the total variation along the interval; see the Appendix for an analysis of the residual errors. Note that this data implies that fcc and hcp the have the extremal values of $\bar{\Lambda}$ among the Barlow packings, at least among the three classes considered here.

\section{Conclusions and Discussion\label{conclusion}}

We characterized the large-scale structure and hyperuniformity of Barlow packings. We demonstrated that all of the Barlow packings, disordered or periodic, are stealthy with a common lower bound for $K$ of $K = 4\pi/\sqrt{3}$. This stealthiness property implies that Barlow packings are Class I hyperuniform, meaning that they can be ranked by the hyperuniformity order metric $\bar{\Lambda}$. We applied this order metric to three classes of Barlow packings, two of which possess certain degrees of stacking disorder, while the third is the small-period Barlow codes. For all of these classes, we found that the data is well-approximated by the simple model of Eq. (\ref{linearmodel}).

There are several noteworthy findings. The first is the computation of a lower bound on $K$ for all Barlow packings, regardless of the degree of stacking disorder. However, we have not shown that this lower bound is necessarily realized. Note that while the local density of the Barlow packings have an intrinsic degree of long-range order in the form of an underlying hexagonal lattice (triangular lattice layers stacked vertically, with no offset), the existence of such an underlying lattice in direct space is typically not sufficient to guarantee that the structure factor has a gap. Rather, the underlying lattice only guarantees periodicity of the structure factor. Thus, the preconditions of the ``stealthy stacking" theorem are qualitatively distinct from the existence of an underlying (Bravais) lattice. In addition, the stacking geometry of the Barlow packings introduces an anisotropic character, as is often the case for configurations that satisfy the stealthy stacking theorem \cite{classical6}. This anisotropy is a fundamental aspect of the geometry of the Barlow packings, however, current research into order metrics such as $\bar{\Lambda}$ have largely focused on those which discard information about anisotropy. It is an open problem to define order metrics that leverage the peculiar relations between direct space and Fourier space in the case of stacking disorder. In addition, only a few studies have been done on fundamental anisotropic characteristics of hyperuniform systems \cite{hypgen,windowshape}.

Another striking finding is that the hyperuniformity order metric $\bar{\Lambda}$, a large-scale characteristic, is largely determined by the linear model in Eq. (\ref{linearmodel}). This model shows that $\alpha$, which is a local property, determines $\bar{\Lambda}$, at least among the Barlow packings. Thus, $\bar{\Lambda}$ for a complex, large-period stacking can be estimated by a statistic that only involves three layers. The case of stacking disordered packings shows that we can expect that the local character to remain for some types of disorder. It is a topic for future research to determine the boundaries of this local character, in particular, whether similar results hold for systems without an underlying lattice. Considered together, these findings raise fundamental questions about the nature of anisotropic stacking disorder, and provide a motivation for the creation of anisotropic order metrics and the study of the hyperuniformity properties of anisotropic systems. 

\begin{figure}[t]
    \includegraphics[width=0.4\textwidth]{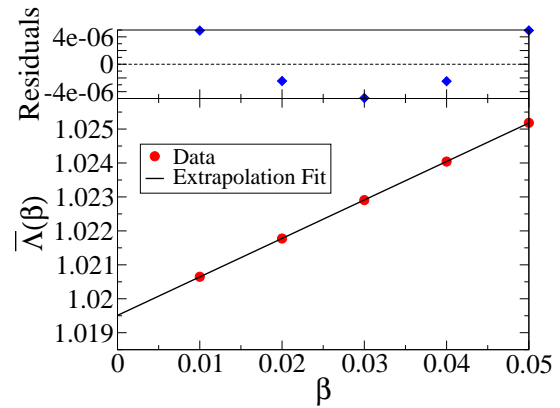}
    \caption{We show a representative fit for the uncorrelated cluster model with $\alpha = 0.5$. Notice how in the residual errors of the fit, there is a quadratic dependence. This implies that we underestimate the value of $\bar{\Lambda}$ with a linear fit.\label{fitexample}}
\end{figure}

\begin{figure*}[t]
    \subfloat[][]{
        \includegraphics[width=0.49\textwidth]{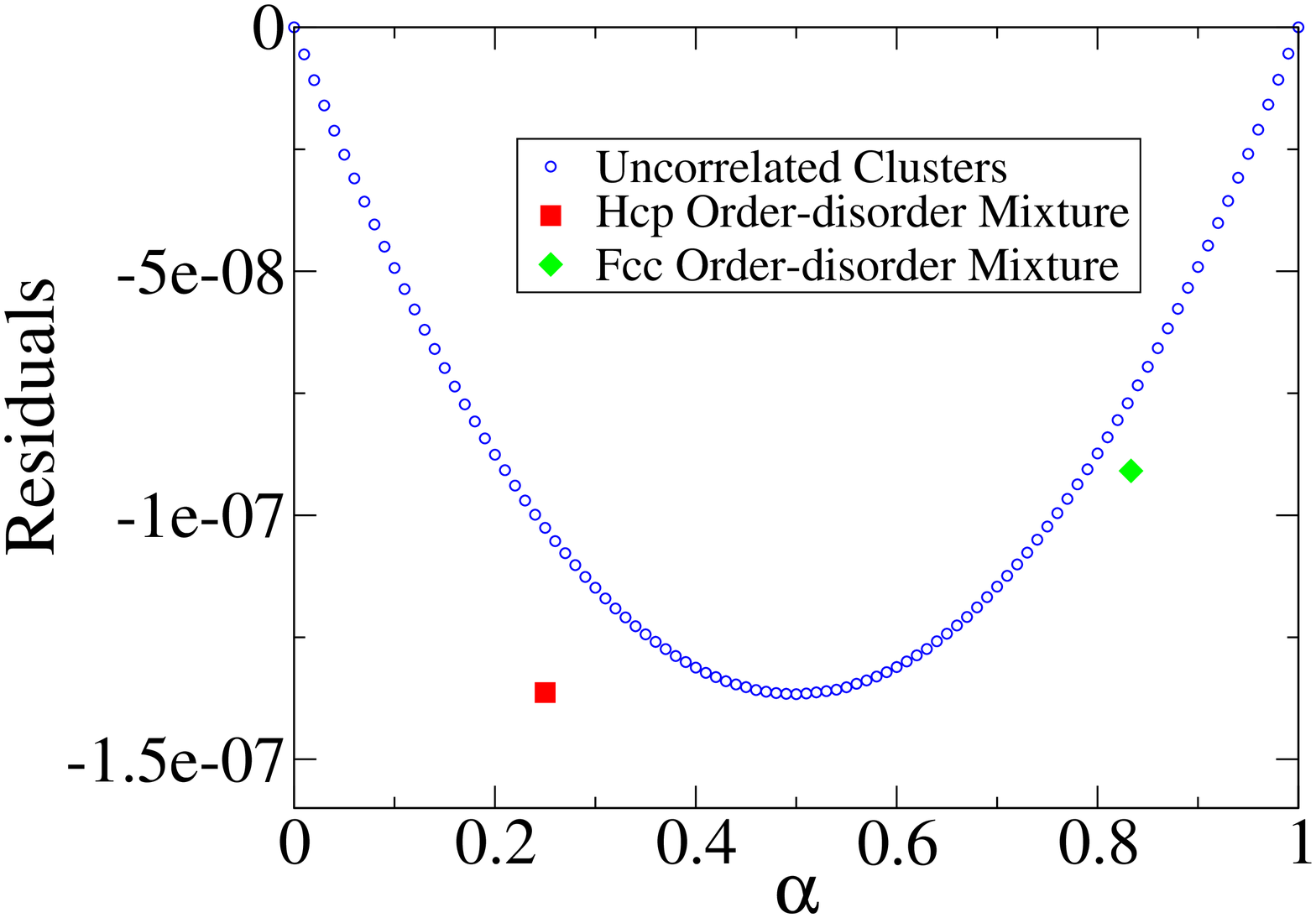}
    }%
    \subfloat[][]{
        \includegraphics[width=0.49\textwidth]{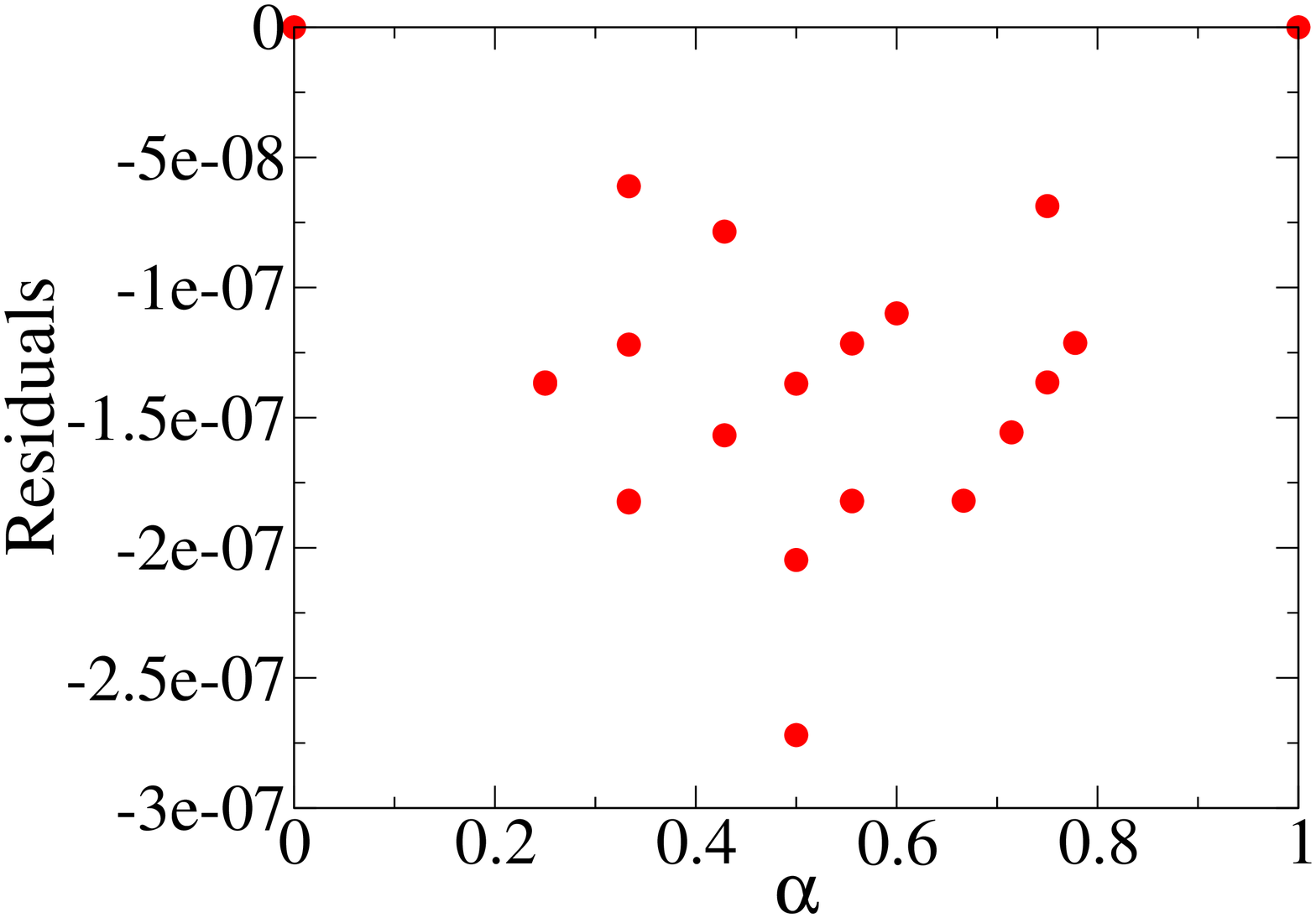}
    }
    \caption{The residual errors between the model (\ref{linearmodel}) and the computed values of $\bar{\Lambda}$. Panel (a) gives the stacking disordered data, with the uncorrelated cluster model in blue, the hcp order-disorder mixture in red, and the fcc order-disorder mixture in green. Panel (b) gives the periodic Barlow data. Note that there are 22 data points in panel (b), but the difference between some points is indiscernibly small on the scale of the plot. The values of $\bar{\Lambda}$ needed for the model (\ref{linearmodel}) are determined by the endpoints of each plot.\label{residuals}}
\end{figure*}

\begin{acknowledgments}
    The authors thank Jaeuk Kim for very helpful discussions and assistance in designing the appearance of Figure \ref{fitexample}, Duyu Chen for very helpful discussions, and Michael Klatt for providing very helpful discussions, Figure \ref{fccvhcp}, and other technical support. This work was supported in part by the National Science Foundation under Grant No. DMR-1714722.
\end{acknowledgments}

\appendix*
\section{Numerical Analysis of the Computation of $\bar{\Lambda}$}

In this Appendix, we analyze a representative extrapolation of $\bar{\Lambda}$ and the residual errors (i.e., the data minus the fitted function evaluated at the corresponding domain point) of the linear model (\ref{linearmodel}). Figure \ref{fitexample} gives an example fit, which is taken from the theta series calculations for the uncorrelated cluster model with $\alpha = 0.5.$ The line in the figure is a linear fit. Notice how the residual errors of this fit are quadratic, implying that we are underestimating $\bar{\Lambda}.$ The true value will be on order $10^{-5}$ higher for this example.

Figure \ref{residuals} gives the residual errors to the model and data plotted in Fig. \ref{data}. It is not clear whether the deviations shown are a result of numerical errors in the computation, or whether they represent the true deviations from linearity, but they are on the order of 1000 times smaller than the total variation in $\bar{\Lambda}$. To test whether the finite value of $\beta$ was likely to play a role in the origin of these deviations, we ran the calculations for the periodic Barlow codes up to 6 letters long using a $\beta =$ $0.01$-$0.05$ extrapolation, but found nearly the same errors. This suggests that the finite value of $\beta$ is not likely to be the cause of the deviation. However, it should be noted that our calculated precision in $\bar{\Lambda}$ is still on the order of the deviations, meaning that further work is needed to draw definitive conclusions on the true nature of the deviations.

\end{document}